\documentclass[cits]{PoS} 
\usepackage{amsmath,bm,booktabs,cite,color,dcolumn,slashed}
\newcolumntype{d}[1]{D{.}{.}{#1}}

\title{Bethe--Salpeter-Motivated Modelling of Pseudo-Goldstone
Pseudoscalar Mesons}\ShortTitle{Bethe--Salpeter-Motivated
Modelling of Pseudo-Goldstone Pseudoscalar Mesons}\author
{\speaker{Wolfgang Lucha}\\Institute for High Energy Physics,
Austrian Academy of Sciences, Nikolsdorfergasse 18,\\A-1050
Vienna, Austria\\E-mail: \email{Wolfgang.Lucha@oeaw.ac.at}}

\abstract{We apply a description of bound states of fermion and
antifermion by means of our approximation to the Bethe--Salpeter
formalism that retains part of the information on relativistic
effects provided by the full fermion propagator to the lightest
pseudoscalar mesons. Therein, the pseudo-Goldstone nature of the
latter quark--antiquark bound states is taken into account by
appropriately formulated effective interactions. Scrutinizing the
predictions of this bound-state approach for meson masses, decay
constants and in-meson condensates by relying on a generalized
Gell-Mann--Oakes--Renner relation shows that the light-quark-mass
values required for agreement are all in the right ballpark.}

\FullConference{XIII Quark Confinement and the Hadron Spectrum ---
Confinement2018\\31 July -- 6 August 2018\\Maynooth University,
Ireland}

\begin{document}\section{Interpretation of the Lightest
Pseudoscalar Mesons as (Pseudo) Goldstone Bosons}Spontaneous
breakdown of the chiral symmetries of quantum chromodynamics (QCD)
implies the presence of massless bosons, identified with the
ground-state pseudoscalar mesons, with masses due to these
symmetries' further, explicit breaking. As a proof of concept, we
study the treatment of such pseudo-Goldstone bosons by a simple
approximation \cite{WLe,WLp,WLpa,WLpb,WL0} to the Bethe--Salpeter
equation \cite{SB}.

\section{Bethe--Salpeter Equation for Fermion--Antifermion States
in Instantaneous Limit}\label{I}Consider some boson bound state
$|B(P)\rangle$ of mass $\widehat M_B$ and momentum $P,$ composed
of a fermion and an antifermion with individual coordinates
$x_{1,2}$, individual momenta $p_{1,2}$, center-of-momentum
coordinate $X$, relative coordinate $x$, total momentum $P$, and
relative momentum $p$, clearly related~by$$x\equiv x_1-x_2\
,\qquad P\equiv p_1+p_2\ ,\qquad P^2=\widehat M_B^2\ .$$The
Bethe--Salpeter formalism describes the bound state $|B(P)\rangle$
by its Bethe--Salpeter amplitude, in momentum space defined, in
terms of the Dirac field operators $\psi_{1,2}(x_{1,2})$ of the
two constituents,~by$$\Phi(p,P)\equiv\exp({\rm i}\,P\,X)\int{\rm
d}^4x\,\exp({\rm i}\,p\,x)\,\langle0|{\rm T}(\psi_1(x_1)\,
\bar\psi_2(x_2))|B(P)\rangle\ .$$This Bethe--Salpeter amplitude
satisfies the homogeneous Bethe--Salpeter equation \cite{SB} that,
in turn, involves both the appropriate interaction kernel and the
propagators of the bound-state constituents. In Lorentz-covariant
settings, the full propagator $S_i(p)$ of any spin-$\frac{1}{2}$
fermion $i$ can be represented in terms of two Lorentz-scalar
functions, \emph{e.g.}, mass $M_i(p^2)$ and wave-function
renormalization~$Z_i(p^2)$, obtained as solutions to the
Dyson--Schwinger equation for the fermion's two-point Green
function:\begin{equation}S_i(p)=\frac{{\rm i}\,Z_i(p^2)}
{\slashed{p}-M_i(p^2)+{\rm i}\,\varepsilon}\ ,\qquad\slashed{p}
\equiv p^\mu\,\gamma_\mu\ ,\qquad\varepsilon\downarrow0\ ,\qquad
i=1,2\ .\label{p}\end{equation}

Some time ago, we devised a (Salpeter-equation-generalizing)
three-dimensional reduction~\cite{WLe} of the Poincar\'e-covariant
Bethe--Salpeter equation, enabled by keeping in fermion
propagators~only terms linear in $p_0$. The latter, together with
the assumption of instantaneity of all interactions~among the
bound-state constituents, suffices to formulate bound-state
equations for Salpeter amplitudes \cite{SE}$$\phi(\bm{p})\propto
\int{\rm d}p_0\,\Phi(p,P)\ .$$In terms of its bound-state
constituents' free energies and projectors onto positive/negative
energies,$$E_i(\bm{p})\equiv\sqrt{\bm{p}^2+M_i^2(\bm{p}^2)}\
,\qquad\Lambda_i^\pm(\bm{p})\equiv\frac{E_i(\bm{p})\pm\gamma_0\,
[\bm{\gamma}\cdot\bm{p}+M_i(\bm{p}^2)]}{2\,E_i(\bm{p})}\ ,$$and
induced interaction kernel $K(\bm{p},\bm{q})$, our
center-of-momentum-frame bound-state equation reads
\begin{align}\phi(\bm{p})=Z_1(\bm{p}^2)\,Z_2(\bm{p}^2)
\int\frac{{\rm d}^3q}{(2\pi)^3}&\left(\frac{\Lambda_1^+(\bm{p})\,
\gamma_0\,[K(\bm{p},\bm{q})\,\phi(\bm{q})]\,\Lambda_2^-(\bm{p})\,
\gamma_0}{\widehat M_B-E_1(\bm{p})-E_2(\bm{p})}\right.\nonumber\\&
\hspace{-.7ex}-\left.\frac{\Lambda_1^-(\bm{p})\,\gamma_0\,
[K(\bm{p},\bm{q})\,\phi(\bm{q})]\,\Lambda_2^+(\bm{p})\,\gamma_0}
{\widehat M_B+E_1(\bm{p})+E_2(\bm{p})}\right).\label{i}\end{align}

The normalization of the Salpeter amplitude $\phi(\bm{p})$ will,
of course, reflect that of the state $|B(P)\rangle$ entering its
definition. For the latter normalization, we adhere to the
relativistically covariant choice$$\langle B(P)|B(P')\rangle=
(2\pi)^3\,2\,P_0\,\delta^{(3)}(\bm{P}-\bm{P}')\ .$$

\pagebreak\noindent Neglecting, for one reason or the other, the
impact of the interaction kernel, this yields the condition
(involving a trace over our bound-state constituents' spinor,
flavour, and colour degrees of freedom)$$\int{\rm d}^3p\,{\rm
Tr}\!\left[\phi^\dag(\bm{p})\,
\frac{\gamma_0\,[\bm{\gamma}\cdot\bm{p}+M_1(\bm{p}^2)]}
{E_1(\bm{p})}\,\phi(\bm{p})\right]=(2\pi)^3\,2\,P_0\ .$$

\section{Application Suggesting Itself: Two Bound-State
Constituents of Identical Flavour}Now, let us adapt our general
instantaneous Bethe--Salpeter formalism \cite{WLe} to just those
physical systems we are actually interested in: bound states of a
quark and an antiquark of precisely the same mass --- tantamount,
as far as only the strong interactions are taken into account, to
bound states of a quark and its own antiquark. For this special
case, we may drop the flavour-related subscript~$i=1,2$ in our
framework, whence the instantaneous Bethe--Salpeter equation
(\ref{i}) simplifies (a little bit)~to\begin{align}\phi(\bm{p})=
Z^2(\bm{p}^2)\int\frac{{\rm d}^3q}{(2\pi)^3}&
\left(\frac{\Lambda^+(\bm{p})\,\gamma_0\,
[K(\bm{p},\bm{q})\,\phi(\bm{q})]\,\Lambda^-(\bm{p})\,\gamma_0}
{\widehat M_B-2\,E(\bm{p})}\right.\nonumber\\&\hspace{-.7ex}
-\left.\frac{\Lambda^-(\bm{p})\,\gamma_0\,[K(\bm{p},\bm{q})\,
\phi(\bm{q})]\,\Lambda^+(\bm{p})\,\gamma_0}{\widehat
M_B+2\,E(\bm{p})}\right).\label{=}\end{align}Clearly, the
spin--parity--charge-conjugation assignment of any pseudoscalar
bound state formed by spin-$\frac{1}{2}$ fermion and
spin-$\frac{1}{2}$ antifermion is given by $J^{P\,C}=0^{-+}$. The
most general Salpeter amplitude $\phi(\bm{p})$ of any such state
may be expanded into only two independent Lorentz-scalar
components, say, $\varphi_{1,2}(\bm{p})$. Recalling its colour
factor, for a bound state of quark and its antiquark this
expansion reads\begin{align*}\phi(\bm{p})=\frac{1}{\sqrt{3}}\left[
\varphi_1(\bm{p})\,
\frac{\gamma_0\,[\bm{\gamma}\cdot\bm{p}+M(\bm{p}^2)]}{E(\bm{p})}+
\varphi_2(\bm{p})\right]\gamma_5\ ,&\\4\int{\rm d}^3p\,
[\varphi_1^\ast(\bm{p})\,\varphi_2(\bm{p})
+\varphi_2^\ast(\bm{p})\,\varphi_1(\bm{p})]=(2\pi)^3\,2\,P_0\
.&\end{align*}

At that stage, the only element still lacking is the
Bethe--Salpeter kernel $K(\bm{p},\bm{q})$, with regard~to its
Dirac structure and its dependence on the momenta $\bm{p}$ and
$\bm{q}$. We tackle this problem in two steps.

\subsection{Dirac Structure of the Bethe--Salpeter Interaction
Kernel by Sticking to Fierz Invariance}We base the determination
of the kernel $K(\bm{p},\bm{q})$ on our trust in Fierz symmetries
and rely for its Dirac structure on a linear combination
corresponding to an eigenstate under Fierz transformations:
\begin{equation}K(\bm{p},\bm{q})\,\phi(\bm{q})\propto
V(\bm{p},\bm{q})\left[\gamma_\mu\,\phi(\bm{q})\,\gamma^\mu
+\gamma_5\,\phi(\bm{q})\,\gamma_5-\phi(\bm{q})\right].\label{FI}
\end{equation}Accordingly, all underlying effective interactions are
subsumed by a single Lorentz-scalar potential function,
$V(\bm{p},\bm{q})$. Assuming the latter to be of convolution type
and to be compatible with spherical symmetry, that is,
$V(\bm{p},\bm{q})=V((\bm{p}-\bm{q})^2)$, allows us to split off
all reference to angular variables and to reduce our bound-state
equation (\ref{=}) to a system of equations for the radial factors
$\varphi_{1,2}(p)$ of the independent components
$\varphi_{1,2}(\bm{p})$, depending on the moduli
$p\equiv|\bm{p}|,q\equiv|\bm{q}|$ of the momenta $\bm{p}$ and
$\bm{q}$, with all interactions encoded by a yet to be found
configuration-space central potential~$V(r)$,~$r\equiv|\bm{x}|$:
\begin{subequations}\begin{align}&E(p)\,\varphi_2(p)+
\frac{2\,Z^2(p^2)}{\pi\,p}\int\limits_0^\infty{\rm d}q\,q\,{\rm
d}r\sin(p\,r)\sin(q\,r)\,V(r)\,\varphi_2(q)=\frac{\widehat M_B}{2}
\,\varphi_1(p)\ ,\label{ie}\\&E(p)\,\varphi_1(p)=\frac{\widehat
M_B}{2}\,\varphi_2(p)\ .\label{ae}\end{align}\label{e}
\end{subequations}

\subsection{Momentum Dependence of our Bethe--Salpeter Interaction
Kernel by Utilizing Inversion}The two (in general, coupled)
Eqs.~(\ref{e}) constitute an eigenvalue problem, with the
bound-state masses $\widehat M_B$ as eigenvalues, for bound states
specified, in momentum-space representation, by the set of radial
wave functions $\varphi_{1,2}(p)$. For \emph{vanishing\/}
eigenvalue, that is, for $\widehat M_B=0$, Eqs.~(\ref{e})
decouple: Eq.~(\ref{ae}) forces $\varphi_1(p)$ to vanish,
\emph{i.e.}, $\varphi_1(\bm{p})=0$. Thus, the corresponding
Salpeter~amplitude reads\begin{equation}\phi(\bm{p})=
\frac{1}{\sqrt{3}}\,\varphi_2(\bm{p})\,\gamma_5\ .\label{0M}
\end{equation}In configuration-space representation, denoting the
free term by $T(r)$, Eq.~(\ref{ie}) then simplifies to a relation
enabling us \cite{WLi,WLia} to find the potential in action,
$V(r)$, provided we know one solution~$\varphi_2(r)$:
\begin{equation}T(r)+V(r)\,\varphi_2(r)=0\qquad\Longrightarrow
\qquad V(r)=-\frac{T(r)}{\varphi_2(r)}\ .\label{V}\end{equation}

In order to get hold of, at least, one of the desired solutions,
we exploit the relationship between the full quark propagator
$S(p)$ --- obtainable as solution to the quark Dyson--Schwinger
equation --- and the Bethe--Salpeter amplitude $\Phi(p,0)$ of
(flavour-nonsinglet) pseudoscalar mesons arising from the
(renormalized) axial-vector Ward--Takahashi identity of QCD in the
chiral limit \cite{MRT}:~the~sought relationship (in its
Euclidean-space formulation indicated by underlined quantities)
reads \cite{WLc,WLca,WLcb,WLp,WLr}\begin{equation}
\Phi(\underline{k},0)\propto
\frac{Z(\underline{k}^2)\,M(\underline{k}^2)}
{\underline{k}^2+M^2(\underline{k}^2)}\,\underline{\gamma}_5+
\mbox{subleading contributions}\ .\label{rs}\end{equation}

\begin{figure}[hbt]\begin{center}\begin{tabular}{cc}
\includegraphics[scale=2.15485]{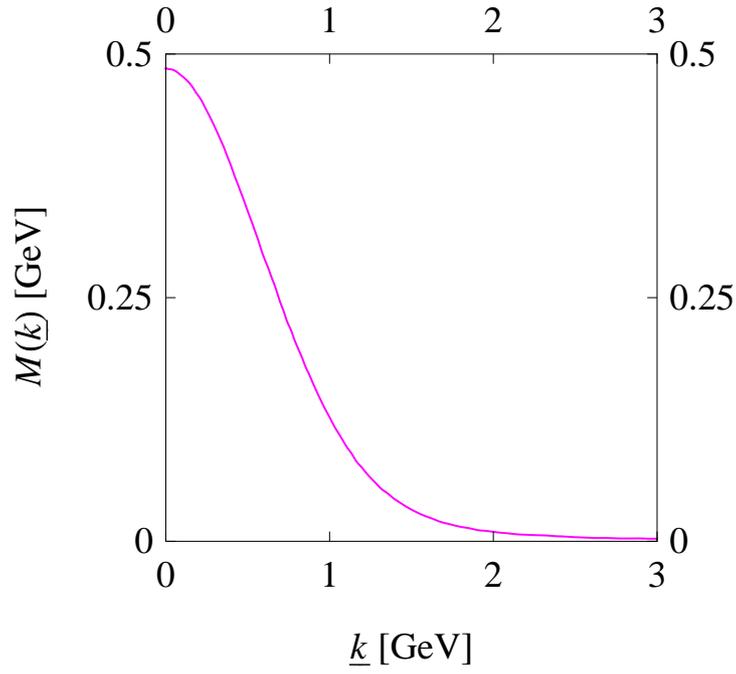}\\[1ex](a)\\[2ex]
\includegraphics[scale=2.15485]{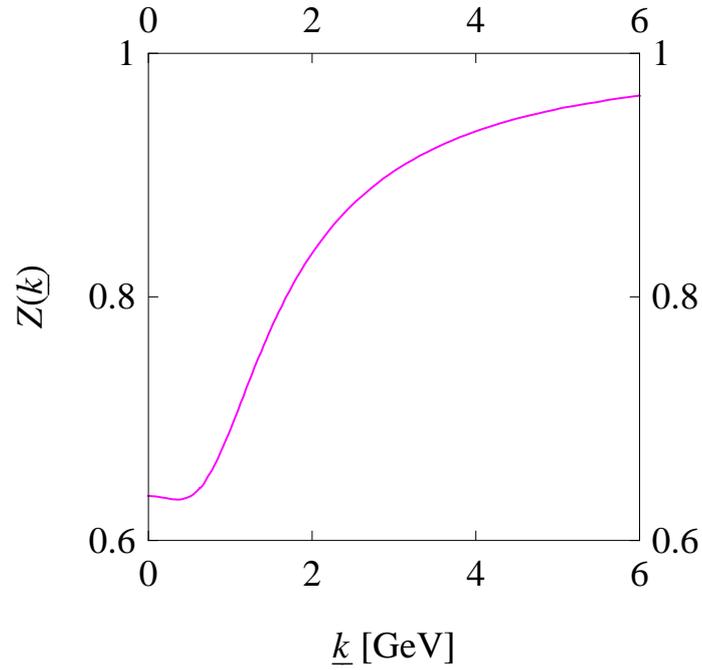}\\[1ex](b)\end{tabular}
\caption{Dependence, on the Euclidean momentum $\underline{k}$, of
both (a) mass function $M(\underline{k})$ and (b) wave-function
renormalization function $Z(\underline{k})$ entering in the quark
propagator (\protect\ref{p}) in the chiral limit,
extracted~and~redrawn \cite{WLp,WLpb} from Fig.~1 of
Ref.~\cite{PM}, emerging as solution to the Dyson--Schwinger
equation for the quark~two-point Green function imitating the
impact of truncated Dyson--Schwinger equations by QCD-inspired
ans\"atze \cite{MT}.\label{MZ}}\end{center}\end{figure}

\begin{figure}[hbt]\begin{center}\begin{tabular}{cc}
\includegraphics[scale=2.01912]{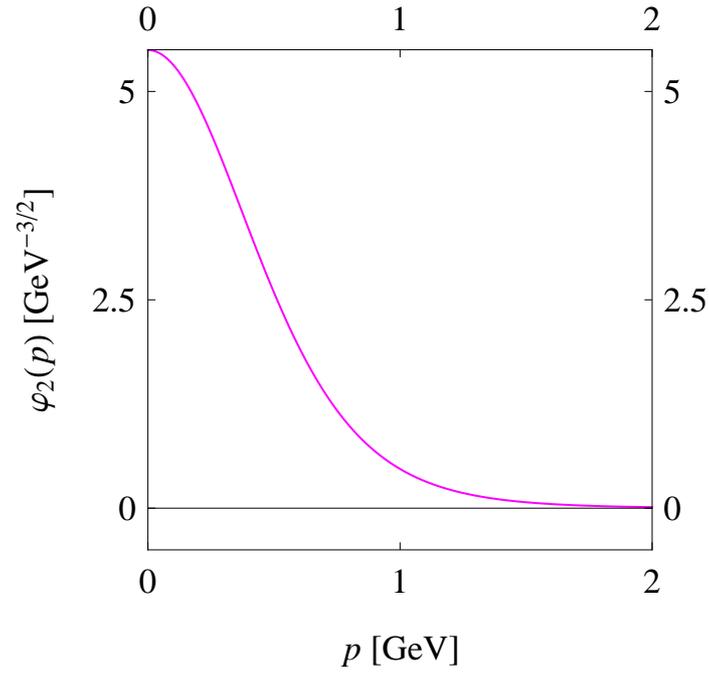}\\[1ex](a)\\[2ex]
\includegraphics[scale=2.01912]{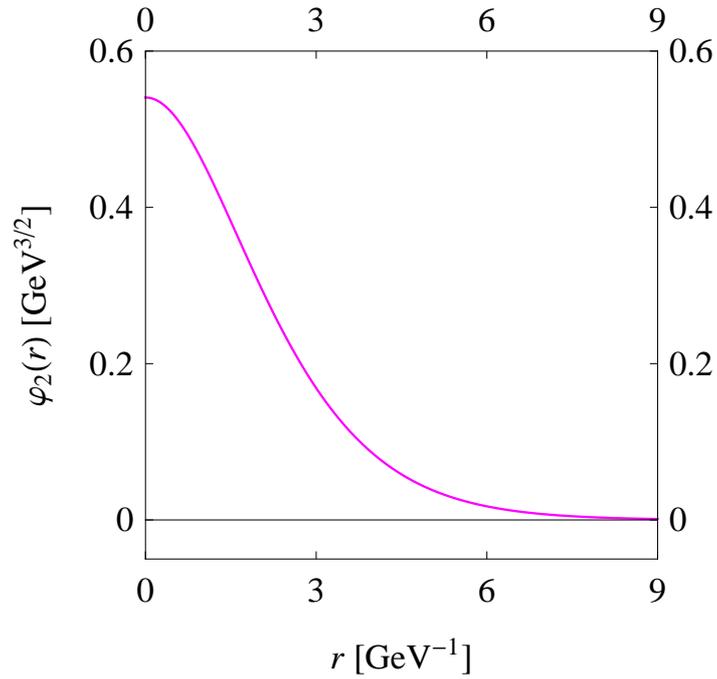}\\[1ex](b)\end{tabular}
\caption{Nonvanishing radial component function determining the
Salpeter amplitude (\protect\ref{0M}) of any massless pseudoscalar
quark--antiquark bound state governed by our instantaneous
Bethe--Salpeter equation (\protect\ref{=}) with
\emph{Fierz-symmetric\/} Dirac structure (\protect\ref{FI}),
extracted from the quark propagator functions $M(\underline{k})$
and $Z(\underline{k})$ of Fig.~{\protect\ref{MZ}}, shown in both
(a) momentum-space representation, $\varphi_2(p)$, and (b)
configuration-space representation, $\varphi_2(r)$.\label{R}}
\end{center}\end{figure}

Just for the sake of illustration, let us follow the path sketched
above by starting from a solution for the chiral-quark propagator
found in Ref.~\cite{MT} on the basis of a particular QCD-motivated
ansatz for the \emph{effective\/} interactions entering in the
quark Dyson--Schwinger equation: the conversion of~the propagator
functions $M(\underline{k})$ and $Z(\underline{k})$ redrawn in
Fig.~\ref{MZ}, by means of Eq.~(\ref{rs}), to the massless-meson
Salpeter amplitude of Fig.~\ref{R} entails, via the inversion
(\ref{V}), the interquark potential plotted in Fig.~\ref{P}.

\section{Basic Pseudoscalar-Meson Features in a
Gell-Mann--Oakes--Renner-type Relation}With the explicit behaviour
of the effective interquark potential $V(r)$ at our disposal, we
are in a position to embark on the intended simplified description
of meson properties: for $\widehat M_B\ne0,$ inserting any of
Eqs.~(\ref{e}) into the other, takes us to a single eigenvalue
equation for eigenvalues~${\widehat M_B}^2$
\cite{WLs,WLp,WLpa,WLpb,WL0}, which can be easily solved by
expanding its solutions over suitable bases in function space
\cite{WLlb,WLst,WLua,WLnv,WLtwr,WLws,WLh,WLy}.

Matching residues of pseudoscalar-meson poles in the axial-vector
Ward--Takahashi identity of QCD gives a
Gell-Mann--Oakes--Renner-resembling \cite{GOR} relation \cite{MRT}
linking, besides meson mass $\widehat M_B$ and two quark masses,
both \emph{decay constant\/} $f_B$ and \emph{in-hadron
condensate\/} ${\mathbb C}_B$ of the pseudoscalar bound state
$|B(P)\rangle$, defined, in terms of quark fields $\psi_f(x)$
(exhibiting the flavour index $f=1,2$),~by\begin{align*}\langle0|
{:\!\bar\psi_1(0)\,\gamma_\mu\,\gamma_5\,\psi_2(0)\!:}|B(P)\rangle
={\rm i}\,f_B\,P_\mu\qquad\Longrightarrow\qquad f_B&\propto
\int{\rm d}^3p\,{\rm Tr}[\gamma_0\,\gamma_5\,\phi(\bm{p})]\ ,\\
\langle0|{:\!\bar\psi_1(0)\,\gamma_5\,\psi_2(0)\!:}|B(P)\rangle
\equiv{\mathbb C}_B&\propto\int{\rm d}^3p\,{\rm
Tr}[\gamma_5\,\phi(\bm{p})]\ .\end{align*}Sticking still to the
idealized case of bound-state constituents of equal mass $m$, this
relation becomes $$f_B\,\widehat M_B^2=2\,m\,{\mathbb C}_B\ .$$

\clearpage

\begin{figure}[hbt]\begin{center}
\includegraphics[scale=2.2004]{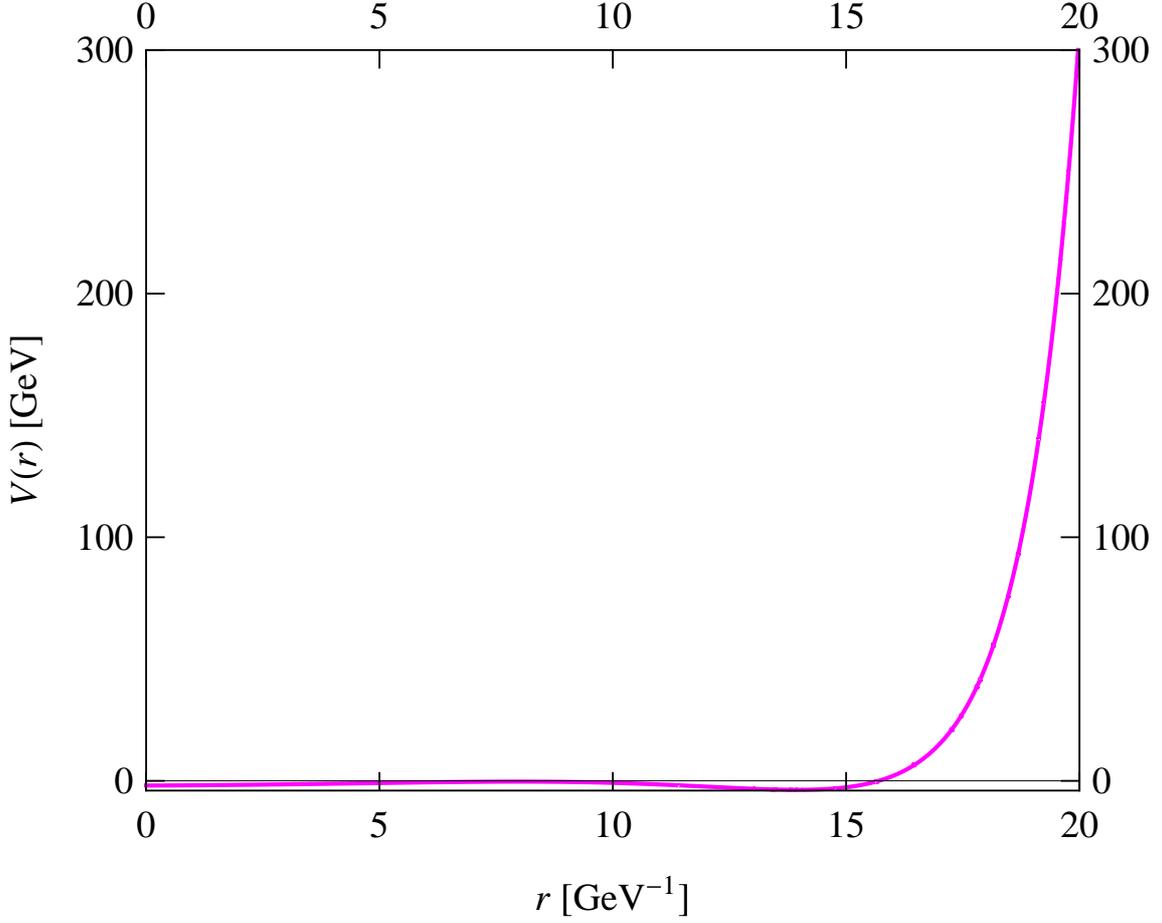}\caption{Spherically
symmetric effective quark--antiquark interaction potential $V(r)$
that upon insertion into our instantaneous Bethe--Salpeter
equation (\protect\ref{e}) trimmed to describe pseudoscalar mesons
reproduces,~as the ground-state solution to that bound-state
problem, the (starting-point) Salpeter component depicted in
Fig.~{\protect\ref{R}}. Its near flatness close to the origin and
steep rise to infinity make $V(r)$ reminiscent of a
\emph{smoothed\/} square~well.\label{P}}\end{center}\end{figure}

Table~\ref{F} presents the very satisfactory outcomes of
implementation of the potential of Fig.~\ref{P}~into our
bound-state approach (see, \emph{e.g.}, Refs.~\cite{GRHKL,WL@,C13}
for corresponding recent Bethe--Salpeter results).

\begin{table}[hbt]
\centering\caption{Gell-Mann--Oakes--Renner-required features of
pseudoscalar quark--antiquark bound states (masses $\widehat M_B,$
decay constants $f_B$ and in-meson condensates ${\mathbb C}_B$)
predicted by Eqs.~(\protect\ref{e}) for the potential $V(r)$ of
Fig.~\protect\ref{P}, and adequate quark masses $m$ compared with
the current-quark values $\overline{m}(\mu)$ in modified minimal
subtraction.}\label{F}\vspace{2ex}\begin{tabular}{rrccd{1.4}c}
\toprule\multicolumn{1}{c}{Constituents}&\multicolumn{1}{c}
{$\widehat M_B$}&$f_B$&${\mathbb C}_B$&\multicolumn{1}{c}{$m$}&\ \
$\overline{m}(2\;\mbox{GeV})$\\&\multicolumn{1}{c}{\ \
$[\mbox{MeV}]$\ \ }&\ \ $[\mbox{MeV}]$\ \ &\ \ $[\mbox{GeV}^2]$\ \
&\multicolumn{1}{c}{\ \ $[\mbox{MeV}]$\ \ }&\ \ $[\mbox{MeV}]$
\cite{PDG}\\\midrule chiral quarks&6.8&151&0.585&0.0059&---\\[1ex]
$u$/$d$ quarks&148.6&155&0.598&2.85& $3.5^{+0.5}_{-0.2}$\\[1ex]$s$
quarks&620.7&211&0.799&51.0& $95^{+9}_{-3}$\\\bottomrule
\end{tabular}\end{table}


\begin{thebibliography}{30}
\bibitem{WLe}W.~Lucha and F.~F.~Sch\"oberl, J.~Phys.~G \textbf{31}
(2005) 1133, arXiv:hep-th/0507281.
\bibitem{WLp}W.~Lucha and F.~F.~Sch\"oberl, Int.~J.~Mod.~Phys.~A
\textbf{31} (2016) 1650202, arXiv:1606.04781 [hep-ph].
\bibitem{WLpa}W.~Lucha, EPJ Web Conf.~\textbf{129} (2016) 00047,
arXiv:1607.02426 [hep-ph].
\bibitem{WLpb}W.~Lucha, EPJ Web Conf.~\textbf{137} (2017) 13009,
arXiv:1609.01474 [hep-ph].
\bibitem{WL0}W.~Lucha, preprint HEPHY-PUB 1000/18 (2018),
arXiv:1807.06245 [hep-ph].
\bibitem{SB}E.~E.~Salpeter and H.~A.~Bethe, Phys.~Rev.~\textbf{84}
(1951) 1232.
\bibitem{SE}E.~E.~Salpeter, Phys.~Rev.~\textbf{87} (1952) 328.
\bibitem{WLi}W.~Lucha and F.~F.~Sch\"oberl, Phys.~Rev.~D
\textbf{87} (2013) 016009, arXiv:1211.4716 [hep-ph].
\bibitem{WLia}W.~Lucha, Proc.~Sci., EPS-HEP 2013 (2013) 007,
arXiv:1308.3130 [hep-ph].
\bibitem{MRT}P.~Maris, C.~D.~Roberts, and P.~C.~Tandy,
Phys.~Lett.~B \textbf{420} (1998) 267, arXiv:nucl-th/9707003.
\bibitem{WLc}W.~Lucha and F.~F.~Sch\"oberl, Phys.~Rev.~D
\textbf{92} (2015) 076005, arXiv:1508.02951 [hep-ph].
\bibitem{WLca}W.~Lucha and F.~F.~Sch\"oberl, Phys.~Rev.~D
\textbf{93} (2016) 056006, arXiv:1602.02356 [hep-ph].
\bibitem{WLcb}W.~Lucha and F.~F.~Sch\"oberl, Phys.~Rev.~D
\textbf{93} (2016) 096005, arXiv:1603.08745 [hep-ph].
\bibitem{WLr}W.~Lucha and F.~F.~Sch\"oberl, Int.~J.~Mod.~Phys.~A
\textbf{33} (2018) 1850047, arXiv:1801.00264 [hep-ph].
\bibitem{MT}P.~Maris and P.~C.~Tandy, Phys.~Rev.~C \textbf{60}
(1999) 055214, arXiv:nucl-th/9905056.
\bibitem{PM}P.~Maris, in \textit{Proceedings of the International
Conference on Quark Confinement and the Hadron Spectrum IV\/},
editors W.~Lucha and K.~Maung Maung (World Scientific, Singapore,
2002), p.~163, arXiv:nucl-th/0009064.
\bibitem{WLs}Z.-F.~Li, W.~Lucha, and F.~F.~Sch\"oberl, Phys.~Rev.~D
\textbf{76} (2007) 125028, arXiv:0707.3202 [hep-ph].
\bibitem{WLlb}W.~Lucha and F.~F.~Sch\"oberl, Phys.~Rev.~A
\textbf{56} (1997) 139, arXiv:hep-ph/9609322.
\bibitem{WLst}W.~Lucha and F.~F.~Sch\"oberl, Int.~J.~Mod.~Phys.~A
\textbf{14} (1999) 2309, arXiv:hep-ph/9812368.
\bibitem{WLua}W.~Lucha, K.~Maung Maung, and F.~F.~Sch\"oberl,
Phys.~Rev.~D \textbf{63} (2001) 056002, arXiv:hep-ph/0009185.
\bibitem{WLnv}W.~Lucha, K.~Maung Maung, and F.~F.~Sch\"oberl,
Phys.~Rev.~D \textbf{64} (2001) 036007, arXiv:hep-ph/0011235.
\bibitem{WLtwr}W.~Lucha and F.~F.~Sch\"oberl, Recent
Res.~Dev.~Phys.~\textbf{5} (2004) 1423, arXiv:hep-ph/0408184.
\bibitem{WLws}W.~Lucha and F.~F.~Sch\"oberl, Int.~J.~Mod.~Phys.~A
\textbf{29} (2014) 1450057, arXiv:1401.5970~[hep-ph].
\bibitem{WLh}W.~Lucha and F.~F.~Sch\"oberl, Int.~J.~Mod.~Phys.~A
\textbf{29} (2014) 1450181, arXiv:1408.4957~[hep-ph].
\bibitem{WLy}W.~Lucha and F.~F.~Sch\"oberl, Int.~J.~Mod.~Phys.~A
\textbf{29} (2014) 1450195, arXiv:1410.5241~[hep-ph].
\bibitem{GOR}M.~Gell-Mann, R.~J.~Oakes, and B.~Renner,
Phys.~Rev.~\textbf{175} (1968) 2195.
\bibitem{PDG}Particle Data Group (M.~Tanabashi \textit{et al.}),
Phys.~Rev.~D {\bf 98} (2018) 030001.
\bibitem{GRHKL}T.~Hilger, M.~G\'omez-Rocha, A.~Krassnigg, and
W.~Lucha, Eur.~Phys.~J.~A {\bf 53} (2017) 213, arXiv:1702.06262
[hep-ph].
\bibitem{WL@}T.~Hilger, M.~G\'omez-Rocha, A.~Krassnigg, and
W.~Lucha, preprint HEPHY-PUB 1003/18 (2018), arXiv:1807.06245
[hep-ph].
\bibitem{C13}T.~Hilger, M.~G\'omez-Rocha, A.~Krassnigg, and
W.~Lucha, preprint HEPHY-PUB 1008/18 (2018), arXiv:1810.01197
[hep-ph].\end{thebibliography}
\end{document}